\documentclass[twocolumn,aps,prb,superscriptaddress]{revtex4}

\usepackage{graphicx}

\begin{document}

\title{An efficient control of Curie temperature $T_C$ in Ni-Mn-Ga alloys}

\author{V.~V.~Khovailo}
\affiliation{National Institute of Advanced Industrial Science and
Technology, Tohoku Center, Sendai 983--8551, Japan}

\author{V.~A.~Chernenko}
\affiliation{Institute of Magnetism, National Academy of Sciences
of Ukraine, Kiev 03142, Ukraine}

\author{A.~A.~Cherechukin}
\affiliation{Institute of UHF Semiconductor Electronics of RAS,
Moscow 117105, Russia}

\author{T.~Takagi}
\affiliation{Institute of Fluid Science, Tohoku University, Sendai
980--8577, Japan}

\author{T.~Abe}
\affiliation{National Institute of Advanced Industrial Science and
Technology, Tohoku Center, Sendai 983--8551, Japan}

\begin{abstract}

We have studied the influence of alloying with a fourth element on
the temperature of ferromagnetic ordering $T_C$ in Ni-Mn-Ga
Heusler alloys. It is found that $T_C$ increases or decreases,
depending on the substitution. The increase of $T_C$ is observed
when Ni is substituted by either Fe or Co. On the contrary, the
substitution of Mn for V or Ga for In strongly reduces $T_C$.

\end{abstract}

\maketitle

Ni-Mn-Ga alloys attract a considerable attention due to the
phenomenon of a large magnetic-field-induced strain observed in
the ferromagnetic state at temperatures below martensitic
transformation temperature $M_s$. The practical importance of this
phenomenon has stimulated an intensive study of composition
dependence of the martensitic transformation temperature. The
experimental results obtained have shown that $M_s$ is very
sensitive to the chemical composition and is observed in a wide
temperature interval, ranging from $< 4$~K up to over
600~K~\cite{c-1995}. Contrary to $M_s$, Curie temperature $T_C$ of
Ni-Mn-Ga alloys was found to be less compositional dependent.
Based on the published experimental results it can be concluded
that the highest $T_C \approx 380$~K is observed in the
stoichiometric Ni$_2$MnGa. A decrease of $T_C$ in alloys with a
deficiency in Mn, which possesses a magnetic moment of $\approx 4
\mu_{\mathrm B}$, is due to the dilution of the magnetic
subsystem. For the alloys with Mn excess, it was suggested that
the decrease in $T_C$ is accounted for by antiferromagnetic
coupling of the extra Mn atoms~\cite{e-2002}. Since large
magnetic-field-induced strains are observed only in the
ferromagnetic state, an efficient control of the ferromagnetic
ordering temperature is of importance for realization of this
effect in a large temperature interval. For this aim we studied
the influence of Fe, Co, V, and In on Curie temperature of
Ni-Mn-Ga alloys.

Polycrystalline ingots of Ni-Mn-Ga with addition of Fe, Co, V, and
In were prepared by a conventional arc-melting method in argon
atmosphere. The ingots were homogenized at 1050~K for nine days
and quenched in ice water. Samples for low-field magnetic
measurements were cut from the middle part of the ingots. Curie
temperature was determined as a minimum on the temperature
derivative of magnetization curve, $dM/dT$ (Fig.~1).

\begin{table}
\caption{Nominal composition (in at.\%) and Curie temperature
$T_C$ (in Kelvin) of the alloys studied}
\begin{tabular}{cccc}

\hline

Composition  &  $T_C$ & Composition & $T_C$ \\

\hline

Ni$_{53.5}$Mn$_{21.5}$Ga$_{16}$In$_9$ & 321 &
Ni$_{53.25}$Co$_{0.75}$Mn$_{21}$Ga$_{25}$ & 353 \\

Ni$_{54.5}$Mn$_{20.5}$Ga$_{20}$In$_5$ & 319 &
Ni$_{52.5}$Co$_{1.5}$Mn$_{21}$Ga$_{25}$ & 359 \\

Ni$_{54}$Mn$_{21}$Ga$_{18}$In$_7$ & 310 &
Ni$_{51.75}$Co$_{2.25}$Mn$_{21}$Ga$_{25}$ & 365 \\

Ni$_{54}$Mn$_{18}$V$_3$Ga$_{21}$In$_4$ & 282 &
Ni$_{54}$Fe$_1$Mn$_{20}$Ga$_{25}$ & 346 \\

Ni$_{53.5}$Mn$_{17.5}$V$_{4}$Ga$_{20}$In$_5$ & 279 &
Ni$_{53}$Fe$_2$Mn$_{20}$Ga$_{25}$ & 356 \\

Ni$_{54.5}$Mn$_{15.5}$V$_5$Ga$_{25}$ & 251 &
Ni$_{52}$Fe$_3$Mn$_{20}$Ga$_{25}$ & 371 \\

Ni$_{54}$Mn$_{13}$V$_8$Ga$_{25}$ & 209 &
Ni$_{51}$Fe$_4$Mn$_{20}$Ga$_{25}$ & 392 \\

 \hline

\end{tabular}
\end{table}

Nominal composition and Curie temperature $T_C$ of the samples
studied is listed in Table~1. It is evident from these data that
substitution of Ga for In or Mn for V results in decrease of
$T_C$. Especially pronounced decrease of $T_C$ is observed in the
case when Mn atoms, at which magnetic moment is localized, are
substituted for V atoms. This is accounted for by the dilution of
the magnetic subsystem. In the case of substitution of Ga for In,
the decrease of $T_C$ is presumably due to an increase of crystal
lattice parameter, induced by In which has a larger atomic radius
than Ga.

\begin{figure}
\begin{center}
\includegraphics*[width=5cm]{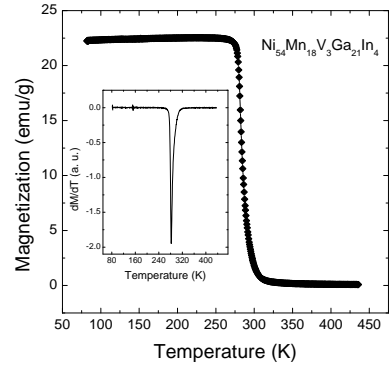}
\end{center}
\caption{Temperature dependence of magnetization $M$ for
Ni$_{54}$Mn$_{18}$V$_3$Ga$_{21}$In$_4$ measured upon heating in a
field $H = 0.01$~T. The inset shows the temperature derivative of
the magnetization, $dM/dT$.}
\end{figure}

Magnetization measurements of two series of samples where Ni was
substituted for Co or Fe showed that Curie temperature increases
with increasing Co (or Fe) content (Table~1). In these alloys Mn
content remains constant (21~at.\% for Co-containing samples and
20~at.\% for Fe-containing samples) and, therefore, the Mn
magnetic subsystem is supposed to be not influenced by these
substitutions. The observed increase of $T_C$ in the Co- and
Fe-containing samples implies that the magnetic properties of
Ni-Mn-Ga should be considered taking into account small magnetic
moments located on Ni atoms~\cite{w-z} and their coupling with
magnetic moments of Mn atoms. For example, the increase in $T_C$
in these alloys can be accounted for by a stronger Co-Mn (Fe-Mn)
exchange interaction as compared to the Ni-Mn one. Besides this
scenario, the enhancement of $T_C$ in Ni-Mn-Ga by addition of Co
or Fe could be explained assuming a noncollinear magnetic
configuration of magnetic moments in the ternary Ni$_2$MnGa
Heusler compound. In a recent theoretical consideration of
magnetic properties of Ni$_2$MnGa and Ni$_2$MnAl Enkovaara
\textit{et. al.}~\cite{e-2003} suggested that lowering of the
total energy at a large spiral wave vector is due to Ni and,
therefore Ni significantly affects $T_C$ in these alloys. The
authors suggested that Curie temperature in Ni$_2$MnGa could be
increased by a substitution of the Ni sites. Note also that this
suggestion is in accordance with an empirical observation that
Curie temperature of Co$_2$MnZ or Cu$_2$MnZ (Z is a non-transition
element) alloys is higher than $T_C$ in corresponding Ni$_2$MnZ
alloys~\cite{w-z}. Our experimental data, however, are not
sufficient to make an unambiguous conclusion about the mechanism
responsible for the increase of $T_C$ and therefore this effect
needs further investigation.

\end{document}